\documentclass[prd,twocolumn,showpacs,preprintnumbers,amsmath,nofootinbib,amssymb,floatfix]{revtex4-1}
\usepackage{graphicx}
\usepackage{subfigure}
\usepackage{amsmath}
\usepackage{amsfonts}
\usepackage{cancel}
\usepackage{multirow}

\begin{document}

\newcommand{\Ima}{\textrm{Im}}
\newcommand{\Rea}{\textrm{Re}}
\newcommand{\mev}{\textrm{ MeV}}
\newcommand{\be}{\begin{equation}}
\newcommand{\ee}{\end{equation}}
\newcommand{\ba}{\begin{eqnarray}}
\newcommand{\ea}{\end{eqnarray}}
\newcommand{\gev}{\textrm{ GeV}}
\newcommand{\nn}{{\nonumber}}
\newcommand{\dtres}{d^{\hspace{0.1mm} 3}\hspace{-0.5mm}}

\renewcommand*{\thesubfigure}{} 




\title{Pseudotensor mesons as three-body resonances}

\author{L.~Roca}
\affiliation{Departamento de F\'{\i}sica. Universidad de Murcia. E-30071, Murcia. Spain}

\date{\today}

 \begin{abstract}
 
We show that the lightest pseudotensor mesons $J^{PC}=2^{-+}$ can be
regarded as molecules made of a pseudoscalar $(P)$ $0^{-+}$ and  a
tensor $2^{++}$ meson, where the latter is itself made of two vector
($V$) mesons. The idea stems from the fact that the  vector-vector
interaction in s-wave and spin 2 is very strong, to the point of
generating the  $2^{++}$ tensor mesons. On the other hand the
interaction  of a pseudoscalar  with a vector meson
in s-wave is also very
strong  and it generates dynamically the lightest axial-vector mesons.
Therefore we expect the $PVV$  interaction to be strongly attractive and
thus able to build up quasibound $PVV$ resonances. We calculate the
three body $PVV$ interaction by  using the fixed center approximation to
the Faddeev equations where the two vectors are clustered forming a
tensor meson. We find clear resonant structures which can be identified
with the  $\pi_2(1670)$, $\eta_2(1645)$ and $K^*_2(1770)$  ($2^{-+}$)
pseudotensor mesons.

\end{abstract}
\maketitle

\section{Introduction\label{sec:1}}

Unveiling the structure and nature of hadrons
is of crucial importance to understand
the strong interaction. 
Several different
components can contribute to the wave functions of the mesonic
resonances besides the simple quark-antiquark state.
In many mesons the quark-antiquark component 
is the dominant one. However,
for some specific mesonic resonances,
other contributions such as glueballs, tetraquarks and meson molecules
can dominate their wave function.
If the meson-meson interaction is attractive 
the meson molecule component may be dominant and the dynamical 
generation of the resonances may be more efficient than the other
Fock space terms.
In the present work, this will be a recurrent idea for many resonances
considered.
In the mesonic sector, important results regarding the molecular
interpretation have
been obtained by 
the unitary extensions of chiral perturbation theory (UChPT or
{\it chiral unitary
approach}). Using as input lowest orders chiral Lagrangians and 
implementing unitarity in coupled channels, many 
resonances are obtained from the meson-meson or meson-baryon
interaction
\cite{Kaiser:1995cy,npa,iam,nsd,Kaiser:1998fi,angels,juanenrique,ollerulf,carmenjuan,hyodo,Jido:2003cb},
which are also usually called dynamically generated resonances.

In particular, it is of special 
interest for the present work the pseudoscalar-vector and
vector-vector unitarized interaction.
In the last few years several 
works~\cite{Lutz:2003fm,Roca:2005nm,Geng:2006yb,Nagahiro:2011jn}  
have reported arguments and evidence for a dynamical nature 
of the lightest
axialvector resonances, implementing variants of the chiral unitary
approach. The
axialvector resonances naturally appear \cite{Lutz:2003fm,Roca:2005nm} as
poles in the scattering matrix of the interaction of pseudoscalar mesons with
vector mesons. Therefore 
most of the low-lying axialvector mesons
can be described by dynamics of a pseudoscalar and a vector meson and
thus can be regarded as molecules made of a pseudoscalar and a vector
meson. 
  
On the other hand, the vector-vector interaction has been recently
studied \cite{Molina:2008jw,gengvec,nagahiro} using the techniques of
the chiral unitary approach, using as input for the vector-vector
potential  the lowest order hidden gauge symmetry Lagrangian
\cite{hidden1,hidden2,hidden3,hidden4}. In \cite{gengvec} eleven
resonant states were found in nine  strangeness-isospin-spin channels.
In particular, and of interest for the present work, the lightest
tensor  $2^{++}$ mesons  $a_2(1320)$, $f_2(1270)$ and $K^*_2(1430)$ were
dynamically generated from the $VV$ interaction in s-wave and spin two
and they where  found to be dominantly molecules made  of $K^*\bar K^*$,
$\rho\rho$ and  $\rho K^*$
 respectively\footnote{Actually, in
Ref.~\cite{gengvec} the pole in  $K^*\bar K^*$ amplitude was
not associated to the $a_2(1320)$ resonance since it was
far from the experimental mass. 
However we discuss in the present work that
the $a_2(1320)$ can indeed be found in this channel with a slight
modification of the only free parameter of Ref.~\cite{gengvec},  as
pointed out in ref~\cite{Molina:2010bk}.}.

From the previous considerations, it is reasonable to expect that a
system  made of a pseudoscalar and two vector mesons ($PVV$) is bound
given the strong attraction between the two vector mesons in s-wave with
parallel spins (which in turn form a tensor meson) and the strong
interaction of the pseudoscalar with the two constituent vector mesons.
These possible bound (or quasibound) states would have $J^{PC}=2^{-+}$
quantum numbers which could correspond to some known (or still
undiscovered) pseudotensor resonances. The main aim of the present work
is to carry out the theoretical study of such  possibility. Indeed, and
bringing forward some results of this work, we will find several $PVV$
resonant structures which may be associated to the $\pi_2(1670)$,
$\eta_2(1645)$ and $K^*_2(1770)$   resonances.

The idea of the existence of three body resonances is of course not new.
However, while much work has been done in the baryonic sector, ({\it e.g.}
\cite{nogami,Ikeda:2007nz,MartinezTorres:2007sr,Jido:2008kp}), less studies
have been devoted to three meson molecules 
\cite{Mennessier:1972bi,MartinezTorres:2008gy,MartinezTorres:2009xb}. 
 The proper analysis of the three body problem, like the one required
in the present work to study the $PVV$ interaction, can be conceptually
tackled by the implementation of the Faddeev equations
\cite{Faddeev:1960su}. However they are very difficult to solve exactly,
hardly ever
possible, and almost always one has to recur to approximations.  For a
recent fresh look into the problem see Ref.~\cite{MartinezTorres:2007sr} for
two meson-one baryon systems and
\cite{MartinezTorres:2008gy,MartinezTorres:2009xb,Albaladejo:2010tj} for
three mesons.

When two of the three particles are bound forming a cluster,  as will be the
case in the present work,  one can use the fixed center approximation (FCA)
to the Faddeev equations
\cite{Chand:1962ec,Barrett:1999cw,Deloff:1999gc,Kamalov:2000iy,Gal:2006cw}. When
applicable, the idea is very simple and considers that one particle collides
against the two particles of the cluster which is not much altered by the
collision, which requires energies close or below threshold
\cite{MartinezTorres:2010ax}. 
Recently, the FCA has been successfully applied in many
three body interactions
\cite{Kamalov:2000iy,Roca:2010tf,YamagataSekihara:2010qk,Xie:2011uw,Xiao:2011rc,Bayar:2011qj}.
In the present work we apply the FCA to the Faddeev equations 
to evaluate the interaction of a
pseudoscalar meson with two vector mesons in spin 2 and s-wave,
where the vector mesons are bound
making up a tensor meson.

\section{Two-body interactions\label{sec:2}}

In the system that we consider in the present work, one pseudoscalar
($P$) and two vector ($V$) mesons, one of the most important ingredients
are the two-body interactions, $VV$ and $PV$. The $VV$ interaction is
needed in order to show that the $VV$ amplitudes in s-wave and spin 2 is
very  attractive, which generate dynamically the lightest $2^{++}$
tensor mesons, and to know to which particular $VV$ channel each tensor
resonance couples most.

On the other hand, 
the $PV$ amplitudes are needed in the FCA equations since we will write the
three-body scattering amplitudes in terms of the two-body interaction
of the
pseudoscalar meson with each of  the two particles in the cluster. We
summarize in what follows the models for the $VV$
\cite{Molina:2008jw,gengvec} and $PV$ \cite{Roca:2005nm} unitarized
interaction properly adapted to the present work.

\subsection{Vector-vector unitarization\label{sec:2.1}}

The model of Refs.~\cite{Molina:2008jw,gengvec} applies
 the ideas of the
chiral unitary approach to the evaluation of the $VV$  scattering amplitudes.
The implementation of unitarity in coupled channels and the exploitation of
the analytic properties of the scattering amplitudes leads to the full
two-body scattering amplitude for a given partial wave, which can be 
written by means of the Bethe-Salpeter equation in coupled channels in the
following way:

\be
t=V+VGt=(1-VG)^{-1}V
\label{eq:bethesalpeter}
\ee
where the kernel $V$ is a matrix containing the elementary
vector-vector transition 
amplitudes  and $G$
is a diagonal matrix with the $l^\textrm{th}$ element, $G_l$, given by the
 loop function for two vector mesons:
\begin{equation}\label{eq:Gsharp}
G_l=i\int\frac{d^4q}{(2\pi)^4}\frac{1}{(P-q)^2-M^2_{l,V1}}\frac{1}{q^2-M^2_{l,V2}},
\end{equation}
where $P$ is the total four-momentum of the $VV$ system and
 $M_{l,V1}$ and $M_{l,V2}$ are the masses of the two vector-mesons
of the corresponding $l^\textrm{th}$ channel.
In the loop function, the widths of the vector mesons are
accounted for by folding Eq.~(\ref{eq:Gsharp}) with their spectral
functions as explained in Ref.~\cite{gengvec}.

\begin{figure}[!h]
\begin{center}
\includegraphics[width=0.8\linewidth]{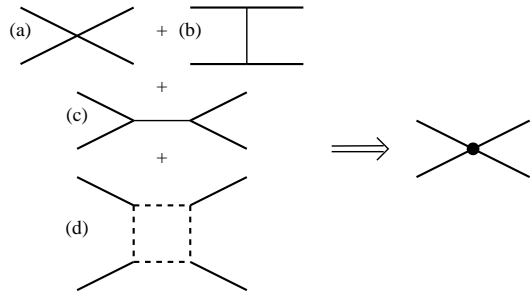}
\caption{Mechanisms contributing to the kernel $V$ (thick dot) of the
Bethe-Salpeter equation, Eq.~(\ref{eq:bethesalpeter}), for 
vector-vector scattering. Solid lines represent vector mesons and dashed
 lines pseudoscalar ones.}
\label{fig:vvdiagram1}
\end{center}
\end{figure}

The mechanisms contributing to the vector-vector potential $V$,
 the kernel of the
Bethe-Salpeter equation (\ref{eq:bethesalpeter}), are depicted in
Fig.~\ref{fig:vvdiagram1}.
The full kernel $V$ is represented by a thick 
dot in Fig.~\ref{fig:vvdiagram1}, to which the mechanisms $(a)$,
$(b)$, $(c)$ and  $(d)$ provide different contributions.
 In this figure
  the solid lines represent vector mesons and the dashed
lines pseudoscalar ones.
For the evaluation of these diagrams 
we need the 4-vectors,
3-vectors and one vector--2-pseudoscalars vertices which
are obtained from the hidden gauge symmetry Lagrangian
\cite{hidden1,hidden2,hidden3,hidden4} for vector mesons.
Explicit expressions
for the Lagrangians and the $V-$matrix elements for the 
different channels
 can be found 
in Refs.~\cite{Molina:2008jw,gengvec,Roca:2010tf}. 
 The dominant contribution to the potential
comes from the contact
term, Fig.~\ref{fig:vvdiagram1}(a),
 and the $t,u$ channel exchange, Fig.~\ref{fig:vvdiagram1}(b). 
The s-channel, Fig.~\ref{fig:vvdiagram1}(c), is very small since it is
basically p-wave. The box diagram, Fig.~\ref{fig:vvdiagram1}(d), is
relevant only for the width of the generated resonance
\cite{Molina:2008jw,gengvec} and it allows the decay into two pseudoscalar
mesons.

The previous formalism can be applied to any possible
strangeness-isospin-spin channel but we are interested in the present work
in the spin 2 channel in s-wave.
In the modulus squared of the different scattering amplitudes, prominent
resonant 
shapes appear (we refer to Ref.~\cite{gengvec} for explicit
plots)  which also correspond to poles in unphysical Riemann sheets
in the complex energy plane, $\sqrt{s}$.
For spin 2 there are three possible
channels. The first one is strangeness 0 and isospin 1 to which 
 $K^* \bar K^*$, $\rho\rho$, $\phi\phi$, $\omega\omega$ and $\omega\phi$
contribute. A pole was found at $\sqrt  s=(1275-1i)$ 
which clearly corresponds to the $f_2(1270)$ resonance.
By evaluating the residues of the scattering amplitudes at the pole
position, the couplings of the dynamically generated $f_2(1270)$ resonance
 to the different channels can be obtained. The 
 coupling to $\rho\rho$ is by far the strongest one \cite{gengvec}.
This is one of the reasons 
why the $f_2(1270)$ resonance can be considered a $\rho\rho$ molecule
or, in other words, a dynamically generated state 
from $\rho\rho$ interaction.
Another of the possible $VV$ spin 2 channels is the strangeness 1,
 isospin 1/2, to which
$K^*\rho$, $K^*\omega$ and $K^*\phi$ contribute in coupled channels.
 The unitarized amplitude in this
case shows up a resonant shape and a pole at $\sqrt{s}=(1431-i1)$ which corresponds
to the $K^*_2(1430)$ resonance. In this case, 
the largest coupling (by a factor 4) 
is to 
$K^*\rho$ channel. Therefore we will consider in the present work the 
$K^*_2(1430)$ as a quasi-bound state of $K^*\rho$ interaction.
Finally, another $VV$ 
channel is possible: strangeness 0 and isospin 1. For
this channel the $K^* \bar K^*$, $\rho\rho$,  $\rho\omega$ and $\rho\phi$
channels are allowed. In Ref.~\cite{gengvec} a pole was found at 
$\sqrt{s}=(1519-i16)$
with the strongest coupling to $K^* \bar K^*$.
The authors of that reference could not clearly assign this pole to 
any experimental $a_2$ resonance.
 However we are going to argue that this channel can produce
the $a_2(1320)$ by doing a fine tunning of the only free parameter in the
model, which is the regulator parameter of the $VV$ loop functions
of Eq.~(\ref{eq:Gsharp}).
The loop function in Eq.~(\ref{eq:Gsharp}) 
needs to be regularized and this
can be accomplished
either
with a three-momentum cutoff or with dimensional regularization. The
equivalence of both methods for meson-meson scattering 
 was shown in Ref.~\cite{iam}.
In Ref.~\cite{gengvec} the regularization method was used 
with subtraction constants, $a$, around a natural value of $-1.65$,
which corresponds to using a three momentum cutoff of
$1$~GeV. With this natural value the bulk of the resonances appear
but a slight fine tuning can be done to agree better with
 the experimental masses of the
$f_2(1270)$ and $K^*_2(1430)$ (see Ref.~\cite{gengvec} 
for the specific values
of the subtraction constants used in the original work),
 but it is worth mentioning that it
only provides a slight modification in the position of the peaks.
We have checked that using the cutoff regularization the peaks of the 
$f_2(1270)$ and $K^*_2(1430)$ are reproduced using three momentum cutoffs
of $875$~MeV for the isospin 0 channel and $972$ for the isospin 1/2
channel. Coming back to the isospin 1 channel we can produce a peak in the
$VV$ amplitude at the position of the $a_2(1320)$
experimental
 mass using a three momentum cutoff of 1590~MeV.
Therefore we can consider the $a_2(1320)$ as a $K^* \bar K^*$ molecule. 
The values of the cutoffs described so far will also play a role in the FCA
equations later on in the evaluation of the tensor form factors.

In summary, in the later evaluation the $PVV$ interaction we
 will regard the $f_2(1270)$ as a cluster made of $\rho\rho$,
  the $K^*_2(1430)$
 as  $K^*\rho$ and the  $a_2(1320)$ as $K^* \bar K^*$.

\subsection{Pseudoscalar-vector unitarization\label{sec:2.2}}

The explicit $PV$ unitarized amplitudes are of crucial importance
in the evaluation of the $PVV$ interaction, since we will need to know the
interaction of the pseudoscalar meson with each of the two vector mesons. 
The $PV$ amplitudes we use in the present work are essentially based on the model of
Ref.~\cite{Roca:2005nm}, where most of the lightest axialvector 
resonances were dynamically generated from the interaction of a vector
and 
a pseudoscalar meson. With the only input of the 
lowest-order chiral Lagrangian and the implementation of 
unitarity in coupled channels the axialvector resonances
manifest themselves as
poles in unphysical Riemann sheets of the $PV$ scattering amplitudes.

Considering the vector mesons as
fields transforming homogeneously under the nonlinear
realization of chiral symmetry \cite{WCCWZ}, 
the interaction of two vector
and two pseudoscalar mesons at lowest order in the pseudoscalar
fields can be obtained from the following interaction Lagrangian 
\cite{Birse:1996hd}: 
\be
\label{eq:lag} {\cal
L}=-\frac{1}{4}\{(\nabla_\mu V_\nu-\nabla_\nu V_\mu)(\nabla^\mu
V^\nu-\nabla_\mu V_\nu) \} \ , \quad 
\ee 
which is invariant under chiral transformations $SU(3)_L\otimes SU(3)_R$.
In Eq.~(\ref{eq:lag}) 
$\nabla_\mu V_\nu=\partial_\mu V_\nu+[\Gamma_\mu,V_\nu]$
is the $SU$(3)-matrix valued covariant derivative, 
with the $SU$(3) connection defined as
$\Gamma_\mu=(u^\dagger \partial_\mu u+u\partial_\mu u^\dagger)/2$,
$u=\exp(P/\sqrt{2} f)$ and $P$ and $V$ are $SU(3)$ matrices containing the
pseudoscalar and vector fields respectively.

From this Lagrangian, the $VP\to VP$ tree level amplitudes 
can be obtained expanding
 Eq.~(\ref{eq:lag})
up to
two vector and two pseudoscalar meson fields:
 \cite{Lutz:2003fm,Roca:2005nm}
\be\label{eq:lag2} 
{\cal L}_{VP}=-\frac{1}{4f^2}\langle[V^\mu,\partial^\nu V_\mu]
[P,\partial_\mu P]\rangle \ , \quad
\ee
where $\langle\rangle$ stands for $SU(3)$ trace.
The 
explicit expression of the potentials, properly projected onto $s$-wave,
is thus
\ba
V_{ij}(s)=&&-\frac{1}{8f^2} C_{ij}
\bigg[3s-(M_i^2+m_i^2+M_j^2+m_j^2)  \nonumber \\
&& -\frac{1}{s}(M_i^2-m_i^2)(M_j^2-m_j^2)\bigg] \ ,
\label{eq:Vtree}
\ea
where  $f=92\mev$ is the pion decay constant,
the index $i(j)$
represents the initial (final) $VP$ state in the isospin
basis and
$M_i(M_j)$ and $m_i(m_j)$ correspond  to the masses of 
the initial (final)
vector mesons and initial (final) pseudoscalar mesons, for which we
use an average value for each isospin multiplet. 
In Eq.~(\ref{eq:Vtree})
we have omitted an $\epsilon_i\cdot \epsilon_j$ term for the
polarization of the vector mesons which factorizes. The explicit values
of the numerical  coefficients, $C_{ij}$, can be found in
Ref.~\cite{Roca:2005nm}.

\begin{figure}[!h]
\begin{center}
\includegraphics[width=0.99\linewidth]{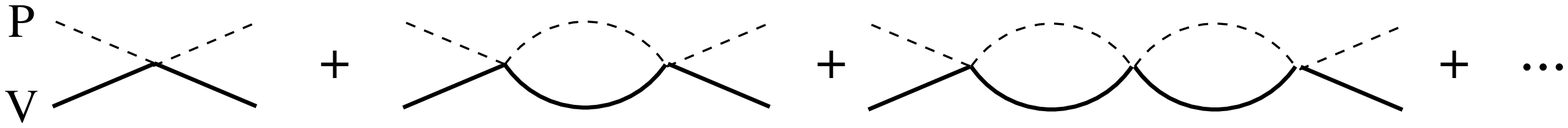}
\caption{Diagrammatic interpretation of
 the unitarization of the $VP\to VP$ amplitude.}
\label{fig:bethe}
\end{center}
\end{figure}

Following the ideas of the chiral unitary approach
the full $PV$ $T$-matrix can now be obtained by unitarizing 
the previous tree level 
amplitudes which in this case leads to the following Bethe-Salpeter
equation:
\be
\label{eq:bethe} 
T=-[1+V G]^{-1}V\\,
\ee
which can be diagrammatically represented by the
resummation series shown in Fig.~\ref{fig:bethe}.

Analogously to Eq.~(\ref{eq:Gsharp}), $G$ 
is a diagonal matrix but now with the $l^{\rm th}$
element, $G_l$, 
given by the loop function of a pseudoscalar and a vector meson,
\be
\label{eq:G}
G_l(P)=i\,\int\frac{d^4 q}{(2\pi)^4} \,
\frac{1}{(P-q)^2-M_l^2+i\epsilon}\,\frac{1}{q^2-m_l^2+i\epsilon}
\ ,
\ee
where $P$ is the total four-momentum, $P^2=s$, of the $VP$ system. 
In order to take into account the width of the vector mesons,
we fold
Eq.~(\ref{eq:G}) by the corresponding vector spectral function 
\cite{Geng:2006yb}.

There are nine different possible channels characterized by their
strangeness ($S$),
isospin ($I$) and $G$-parity ($G$),
but not all them have resonant poles in unphysical Riemann sheets of 
the
complex energy plane, {\it i.e.} do not
generate dynamically resonances.
The channels that manifest resonant poles are
$(S,I,G)=(0,0,+)$, for which $K^*K(+)$ is possible;
$(S,I,G)=(0,0,-)$ for which
 $\phi\eta$, $\omega\eta$,
$\rho\pi$ and $K^*K(-)$ are allowed; 
$(S,I,G)=(0,1,+)$
with
$K^*K(+)$, $\phi\pi$, $\omega\pi$, $\rho\eta$ as allowed channels;
$(S,I,G)=(0,1,-)$
with $\rho\pi$ and $K^*K(-)$ and
$(S,I)=(1,1/2)$ where $\phi K$, $\omega K$,
$\rho K$, $K^*\eta$ and $K^*\pi$ channels are allowed.
In the above paragraph $K^*K(\pm)$ represent the $G$-parity
 eigenstates\footnote{Recall  that the $G$-parity operation can be
defined through its action on an eigenstate of hypercharge
($Y$), isospin  ($I$), and third isospin projection ($I_3$)  as 
$G|Y,I,I_3\rangle=\eta(-1)^{Y/2+I}|-Y,I,I_3\rangle$, with $\eta$ being the
charge conjugation of a neutral non-strange member of the
$SU(3)$ family.} 
$1/\sqrt{2}(|\bar K^* K\rangle\pm|K^*\bar K\rangle)$ with eigenvalues $\pm 1$.

In Ref.~\cite{Roca:2005nm} seven poles were found in the
unphysical Riemann sheets of the unitarized scattering
amplitudes which can be
associated to most of the lightest axial-vector resonances
quoted in the Particle Data Group tables (PDG): $b_1(1235)$, $h_1(1170)$,
$h_1(1380)$, $a_1(1260)$, $f_1(1285)$ and the $K_1(1270)$ resonance.
Actually two poles are present in the unitarized $PV$ amplitude
 for the $K_1(1270)$ resonance. This double
pole structure was studied in Ref.~\cite{Geng:2006yb}.
In Ref.~\cite{Roca:2005nm} all these resonances were obtained using
 a single
value of the parameter needed to regularize the $PV$ loop  function, 
Eq.~(\ref{eq:G}). This parameter was a subtraction constant
 $a=-1.85$ for the dimensional regularization
method or three-momentum cutoff $q_{max}=1$~GeV. However we can now 
fine
tune slightly the subtraction constant or the cutoff to agree better
with the experimental value of the axialvector resonances.
 Furthermore, as
done in Ref.~\cite{Geng:2006yb}, we can also allow that the $f$ constant in
Eq.~(\ref{eq:lag2}) may be $f=115$~MeV instead of $92$~MeV
 in some cases due to the kaon and eta
effects.
Thus, in the present work we use the following values of the subtraction
constant, $a$, and $f$ for the different channels:
for $I=1$,                  $a=-1.95$, $f=92$~MeV;
for $I=0$, $G$-parity $+$,  $a=-1.88$, $f=92$~MeV;
for $I=0$, $G$-parity $-$,  $a=-0.80$, $f=115$~MeV;
and for $I=1/2$,            $a=-1.85$, $f=115$~MeV.
It is important to emphasize that once this slight fine tune of the
regularization parameters is done, the model discussed later in the present work
for the three-body interaction
will have no further freedom.

\begin{figure}[!h]
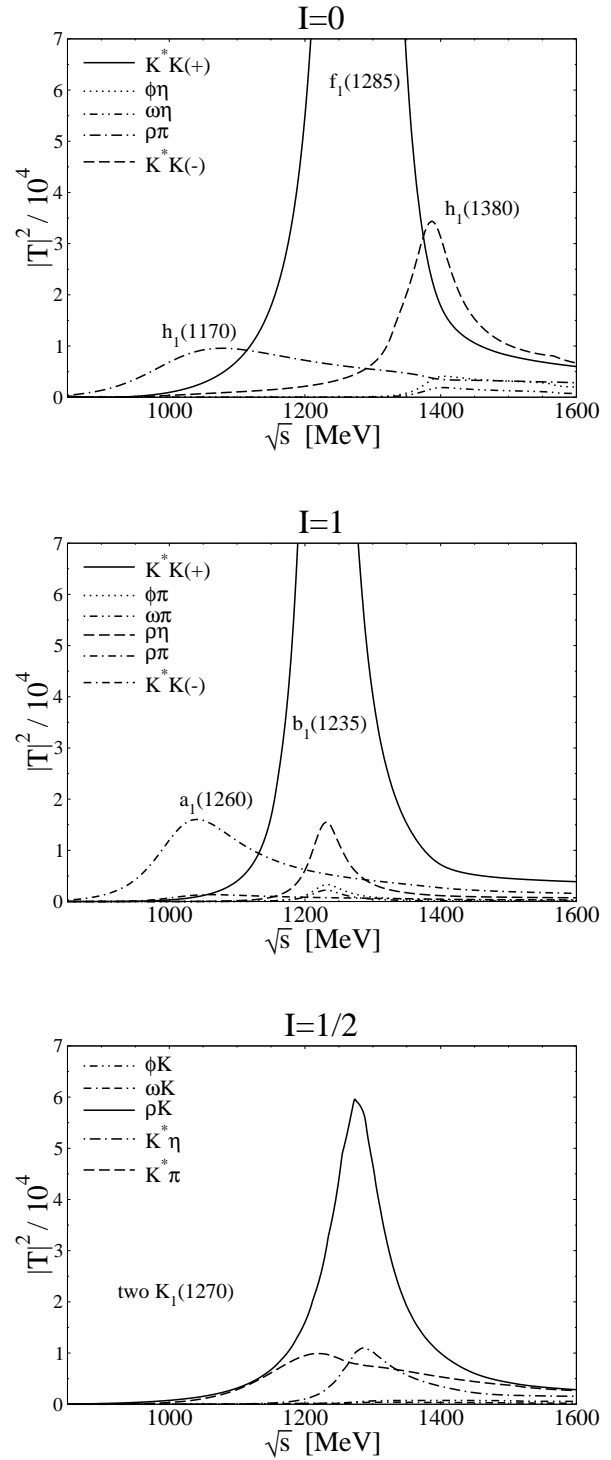

     \centering
      \subfigure[]{
          \includegraphics[width=.9\linewidth]{figure3a.eps}} \\
      \subfigure[]{
          \includegraphics[width=.9\linewidth]{figure3b.eps}} \\
       \subfigure[]{
          \includegraphics[width=.9\linewidth]{figure3c.eps}}
   \caption{ Modulus squared of the pseudoscalar-vector unitarized
scattering amplitudes for the different isospin channels. The 
labels indicate the correspondent experimental particles to which the
resonances dynamically generated are associated.}
     \label{fig:PVamplitudes}
\end{figure}

In Fig.~\ref{fig:PVamplitudes} we show the modulus squared
 of some of the diagonal 
$PV$
unitarized amplitudes. Note that the plots differ slightly 
from those 
in Ref.~\cite{Roca:2005nm} since we have used now slight different values of
the subtraction constants and the loop function with dimensional
regularization is now folded with the vector meson width, which was not done
in Ref.~\cite{Roca:2005nm}. 

It is worth stressing that the unitarized amplitudes provide not only the
masses and widths of the dynamically generated resonances but also the
actual shape of the scattering amplitude. Certainly that includes
whatever background it could
contain which would be also generated thanks to the highly
 non-linear dynamics
involved in the unitarization procedure. Thus, even if there is no
resonance in a
particular channel the method provides the right amplitude for it.

\section{Three-body interaction\label{sec:3}}

In this section we explain the technical details for the
evaluation of the three body interaction between one pseudoscalar particle
and two vector mesons in spin two and s-wave.
As seen in section~\ref{sec:2.1}, whenever we have two vector
mesons with parallel spins they tend to bind very strongly. For instance,
we have seen that two $\rho$ mesons in spin two and s-wave bind very
strongly forming an $f_2(1270)$. This implies a binding energy per $\rho$
particle of about $140$~MeV which is almost 20\% of the $\rho$ meson mass.
 Similar qualitative reasonings can be done for the other tensor
mesons discussed in section \ref{sec:2.1}, 
($K^*_2(1430)$ as a $K^*\rho$ cluster and
 $a_2(1320)$ as $K^* \bar K^*$).
Therefore we can expect that inside the three body $PVV$ system with
spin two the
two vector mesons will be clustered forming a tensor meson. In such
a case, we can apply the fixed center approximation to the Faddeev
equations, which otherwise would be very difficult to solve exactly. As
mentioned in the introduction, the FCA has been proved suited and 
good enough in similar
three body systems when two of the particles tend to cluster together.

For the technicals details we follow closely the steps of
Refs.~\cite{Roca:2010tf,YamagataSekihara:2010qk} with the proper 
adaptations
and modifications to the present case.
In what follows  we are going to explain generically 
the interaction of a particle  
$A$ with a cluster $B$ made 
of two particles, $b_1$ and $b_2$.
For the present work the particle $A$ will represent the pseudoscalar
particle and $b_1$, $b_2$,  the two vector mesons 
that build up the tensor cluster $B$.
Specifically, we study the system $PVV$ with $J^{PC}=2^{-+}$ with three
different possible isospin, $I=1$, $0$, and $1/2$. The $VV$ clusters are the
$2^{++}$ tensor mesons $a_2(1320)$, $f_2(1270)$ and $K^*_2(1430)$ and the
pseudoscalars are $\pi$, $K$ and $\eta$. In the following we 
represent $a_2(1320)$, $f_2(1270)$ and $K^*_2(1430)$ by 
 $a_2$, $f_2$ and $K^*_2$ respectively.
\begin{table}[h]
\begin{center}
\begin{tabular}{|c|c|c|c|c|} 
\hline
\begin{tabular}{c}  total $PVV$\\ isospin\end{tabular} &
\multicolumn{4}{c|}{channels $AB(b_1b_2)$} \\ \hline
$I=1$  & $\pi f_2(\rho \rho)$  & $\eta a_2(K^* \bar K^*)$   &&  \\ \hline
$I=0$ & $\pi a_2(K^* \bar K^*)$  & $\eta f_2(\rho \rho)$ && \\ \hline
$I=\frac{1}{2}$ &  $\pi K^*_2(\rho K^*)$  & $K a_2(K^* \bar K^*)$   & $K
f_2(\rho \rho)$       & $\eta K^*_2(\rho K^*)$ \\ \hline
 \end{tabular}
\end{center}
\caption{Possible pseudoscalar-tensor channels for the
different total $PVV$ isospins. The particles in parenthesis represent the
main vector-vector component of the tensor resonances.
 The $a_2$, $f_2$ and $K^*_2$ symbols
 stand for
  $a_2(1320)$, $f_2(1270)$ and $K^*_2(1430)$ respectively.}
\label{tab:PTchannels}
\end{table}

The allowed channels for the different isospins, taking also into account
the $G$-parity restrictions, are shown in table~\ref{tab:PTchannels}.
These are: for $I=1$, $\pi f_2$ and $\eta a_2$; for
$I=0$, $\pi a_2$ and $\eta f_2$; and for $I=1/2$, $\pi K^*_2$,
$K a_2$, $Kf_2$ and $\eta K^*_2$.
Note that in principle for isospin $I=1$ the $K \bar K^*_2$ and 
$\bar K K^*_2$ are possible with the negative $G$-parity combination
$1/\sqrt{2}|K \bar K^*_2 + \bar K K^*_2\rangle$, however, when doing later 
the
$PV$ interaction to evaluate the three body amplitude, the possible
configurations with these channels
do not respect the total $G$-parity and are thus not allowed.
A similar argument forbids the $1/\sqrt{2}|K \bar K^*_2 - \bar K K^*_2\rangle$
$G$-parity (+) channel in the $I=0$ case.
(This is explained in further detail at the end of the Appendix).

\begin{figure}[!t]
\begin{center}
\includegraphics[width=0.99\linewidth]{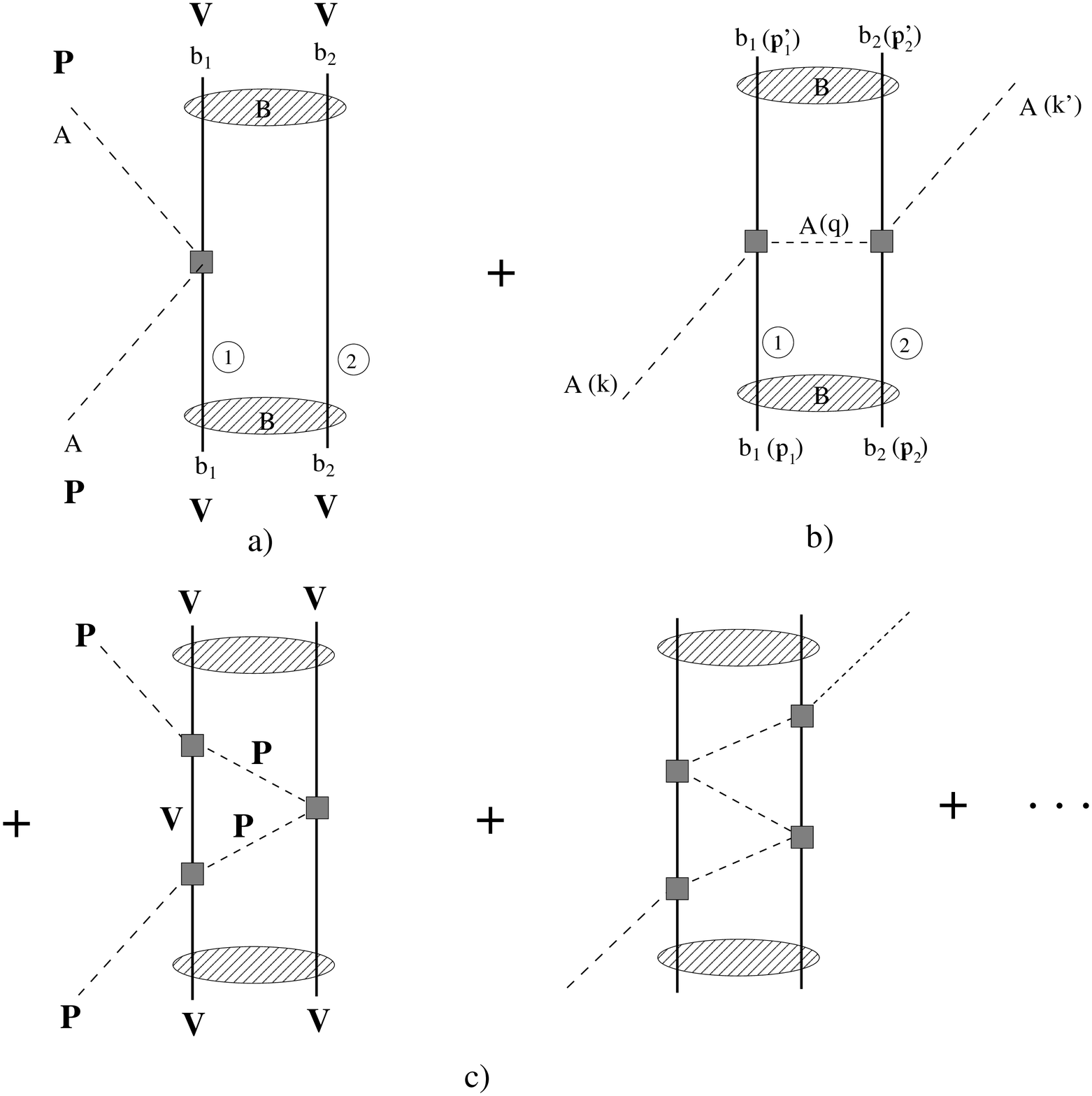}
\caption{Diagrammatic representation of the fixed center approximation
to the Faddeev equations for the interaction of a pseudoscalar particle, 
$A$, with a tensor
particle, $B$, which is a cluster made of two vector mesons,
 $b_1$ and $b_2$.
 Diagrams $a)$ and $b)$ represent
the single and double scattering contributions  respectively and $b)+c)$ the
multiple scattering contribution.}
\label{fig:faddeev}
\end{center}
\end{figure}

The FCA to the Faddeev
equations is represented diagrammatically in Fig.~\ref{fig:faddeev}.
The pseudoscalar particle, $A$, rescatters 
repeatedly
with each of the vector mesons, $b_i$,
which form the tensor resonance cluster, $B$.
The thick squared dots in the figure represent the unitarized $PV$
interaction discussed in section~\ref{sec:2.2}.
 In the figure only the interaction
starting with particle $b_1$ is represented, but an analogous mechanism
 where
particle $A$ starts the interaction against particle $b_2$ must also be
considered. 
Mathematically, the FCA  can be written as a system of coupled equations 
\begin{eqnarray}
T_1&=&t_1+t_1G_0T_2\nonumber\\
T_2&=&t_2+t_2G_0T_1\nonumber\\
T&=&T_1+T_2
\label{eq:faddeev}
\end{eqnarray}
 where  $T_1$, $T_2$, are the two partition functions which 
  sum up to the total scattering matrix, $T$. The $T_i$ amplitudes
accounts for all the diagrams starting with the interaction of the
particle $A$ with particle $b_i$ of the compound  system $B$.
In   Eq.~(\ref{eq:faddeev})
$G_0$ is the Green function for the exchange of a particle $A$ between
the $b_1$ and $b_2$ particles (intermediate dashed lines 
in Fig.~\ref{fig:faddeev})
which expression will be given below, (see Eq.~(\ref{eq:G0})).
 The mechanism in Fig.~\ref{fig:faddeev}a represents the
single-scattering contribution 
($t_1$ in Eq.~(\ref{eq:faddeev}))
and Fig.~\ref{fig:faddeev}b
the double-scattering mechanism (the next contribution:  
$t_1G_0t_2$). 
The addition of Fig.~\ref{fig:faddeev}c represents the full resummation
of mechanisms to get the full $T_1$ partition function in the FCA. 
An analogous figure
starting with the particle $A$ interacting with $b_2$ would account for the 
$T_2$ amplitude.

Note that the FCA equations, Eq.~(\ref{eq:faddeev}), 
are essentially given in 
terms of the two-body pseudoscalar-vector amplitudes,
$t_1$ and $t_2$, for which we used
the unitarized $PV$ amplitudes given in section~\ref{sec:2.2}.
The
 argument of the function $T(s)$ in Eq.~(\ref{eq:faddeev})
 is the total invariant mass energy of the $PVV$ system,
$s$. However the argument of $t_1$ and $t_2$ are $s_1$ and $s_2$, where
$s_i (i=1,~2)$ is the invariant mass of the interacting particle $A$ and
the particle $b_i$ of the $B$ molecule and is given by

\begin{equation}
s_i=m_A^2+m_{b_i}^2+\frac{1}{2 m_B^2}(s-m_A^2-m_B^2)
(m_B^2+m_{b_i}^2-m_{b_{j\ne i}}^2),
\label{eq:si}
\end{equation}
where $m_{A(B)}$ is the mass of the $A(B)$ 
system and $m_{b_i}$ is the mass of each vector meson
 of the $B$ molecule.

The derivation of the expression of the single 
scattering amplitudes, $t_i$, 
in terms of the unitarized two body $PV$
amplitudes of section~\ref{sec:2.2} is explained in detail in 
the Appendix.

Proceeding in an analogous way to the derivation done 
in Refs.~\cite{Roca:2010tf,YamagataSekihara:2010qk},
properly adapted to the
present problem, we can obtain the $S$ matrix for the single scattering
contribution:
\ba
S^{(1)}=S^{(1)}_1+S^{(1)}_2  ,
\ea
with
\ba
&&S^{(1)}_i=-it_{Ab_i} \frac{1}{{\cal V}^2}
\frac{1}{\sqrt{2\omega_{p_i}}}
\frac{1}{\sqrt{2\omega_{p'_i}}}
\frac{1}{\sqrt{2\omega_k}}
\frac{1}{\sqrt{2\omega_{k'}}}\nonumber\\
&&\times(2\pi)^4\,\delta(k+k_{B}-k'-k'_{B})
F_{B,i}\left(\frac{\mu}{m_i}(\vec k - \vec k')\right) .
\label{eq:single}
\ea
where 
$t_{Ab_i}$ are the single scattering amplitudes given in the Appendix,
$\cal V$ represents the volume of a box where we normalize to unity
the plane wave states, 
the momenta are defined in Fig.~\ref{fig:faddeev}b,
 $\omega_p$ represents the on-shell energy of the corresponding
particle with momentum $p$, $k_B$ ($k'_B$)
represents the total momentum of the
initial (final) cluster $B$ and $\mu$ is the reduced mass of the $b_1b_2$
system with masses $m_1$, $m_2$, respectively.
In Eq.~(\ref{eq:single}), $F_{B,i}$ is the form factor 
of the particle $B$ which represents
essentially the Fourier transform of its wave function.
The derivation
of the form factor is similar to the one done in
Refs.~\cite{Roca:2010tf,YamagataSekihara:2010pj},
 where we refer also for
further discussion
and interpretation.
The form factor has to be projected onto s-wave, the one we are
considering in the present work. Hence, $F_{B,i}$ in
 Eq.~(\ref{eq:single}) is replaced by 
\be
F^{(s)}_{B,i}(s)= \frac{1}{2}\int_{-1}^1 d\cos \theta_s\,
F_B(\vec v)
\label{eq:ffactswave}
\ee
where $\vec v\equiv\frac{\mu}{m_i}(\vec k - \vec k')$
with $|\vec v|=\frac{\mu}{m_i} k \sqrt{2(1-\cos \theta_s)}$,
$k=\sqrt{(s-(M_A+M_B)^2)(s-(M_A-M_B)^2})/2\sqrt{s}$
 above 
the $AB$ threshold and zero below.
The $B$ resonance form factor, $F_B(\vec v)$, is 
given by
\ba
F_B(\vec v)&&=\frac{1}{ {\cal N}}
\int_{\substack{|\vec p|<\Lambda\\|\vec p-\vec v|<\Lambda}}
d^3p\,
\frac{1}{m_B-\omega_1(\vec{p})-\omega_2(\vec{p})
+i\frac{\Gamma_1+\Gamma_2}{2}}
\nonumber \\
&&\times \frac{1}{m_B-\omega_1(\vec p-\vec v)
                  -\omega_2(\vec p-\vec v)
		  +i\frac{\Gamma_1+\Gamma_2}{2}},
\label{eq:ffact}
\ea 
\ba
{\cal N}=
\int_{|\vec p|<\Lambda}d^3p
\frac{1}{\left( m_B-
\omega_1(\vec{p})-\omega_2(\vec{p})+i\frac{\Gamma_1+\Gamma_2}{2}
\right)^2}.
\nn
\label{eq:ffactnorm}
\ea 
where $\omega_j(\vec q)=\sqrt{\vec{q}\,^2+m_j^2}$ and
 $\Gamma_j$ is the width
of the $b_j$ particle.
In Eq.~(\ref{eq:ffact}), $\Lambda$ represents a 
three-momentum cutoff 
with a similar physical meaning \cite{Roca:2010tf}
as the three-momentum cutoff of the
vector-vector loop function of
Eq.~(\ref{eq:Gsharp}). For that reason we take for
$B=a_2$, $f_2$ and $K^*_2$ the same values for $\Lambda$
as for the cutoffs mentioned in section \ref{sec:2.1} for the corresponding
channels. The $1/{\cal N}$ factor is introduced in order to normalize to unity
the form factor at zero momentum.

\begin{figure}[!h]
\begin{center}
\includegraphics[width=0.9\linewidth]{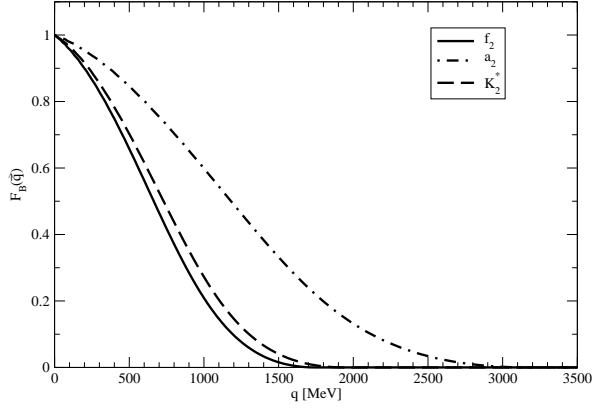}
\caption{Form factors for the $a_2$, $f_2$ and $K^*_2$ tensor mesons.}
\label{fig:ffacts}
\end{center}
\end{figure}

In Fig.~\ref{fig:ffacts} we show the form factors for the 
$a_2$, $f_2$ and $K^*_2$ as a function of the modulus of the
momentum.

The inclusion of the form factor in the single scattering contribution, 
Eq.~(\ref{eq:single}), can be relevant
for energies far above the $AB$ threshold. In the present work we are above
threshold only in the channels where the particle $A$ is a pion. 
We have
checked that, in any case, the numerical effect of this form factor in the
single scattering mechanism is
small. However for the multiple scattering mechanisms the form factor is a
key ingredient as we shall see below.


The next order contribution of the FCA to the Faddeev equations is 
the double scattering mechanism,
which corresponds to Fig.~\ref{fig:faddeev}b.
The $S$-matrix for this
contribution
takes the form
\ba
S^{(2)}=S^{(2)}_1+S^{(2)}_2  ,
\ea
with
\ba
&&S^{(2)}_i=-i(2\pi)^4 \delta(k+k_B-k'-k'_B)\frac{1}{{\cal V}^2}
\frac{1}{\sqrt{2\omega_k}} 
\frac{1}{\sqrt{2\omega_{k'}}}\nn\\
&&\times\frac{1}{\sqrt{2\omega_{p_1}}}
\frac{1}{\sqrt{2\omega_{p'_1}}}
\frac{1}{\sqrt{2\omega_{p_2}}} 
\frac{1}{\sqrt{2\omega_{p'_2}}}\nn\\
&&\times\int \frac{d^3q}{(2\pi)^3} 
F_B\left(\vec q-\frac{\vec k m_{j\ne i}+\vec k' m_i}{m_1+m_2}\right)
\frac{1}{{q^0}^2-\vec{q}\,^2-m_A^2+i\epsilon}\nn\\
&&\times t_{Ab_1} t_{Ab_2}.
\label{eq:finalS2}
\ea
with  $q^0(s;A,B)=(s-m_A^2-m_B^2)/(2m_B)$.
The  term 
$\frac{{\vec k m_{j\ne i}+\vec k' m_i}}{m_1+m_2}$ 
inside the argument of the
form factor is small for bound states, below threshold, and thus it is not
considered in previous works regarding the FCA.
If one considers this term, then $F_B$ must be projected onto s-wave as in 
Eq.~(\ref{eq:ffactswave}). We have checked that the
term $\frac{{\vec k m_{j\ne i}+\vec k' m_i}}{m_1+m_2}$
inside the argument of the form factor has a small numerical effect and
thus we have not considered it in the numerical evaluations for
computational time reasons.

\begin{figure}[!h]
\begin{center}
\includegraphics[width=0.9\linewidth]{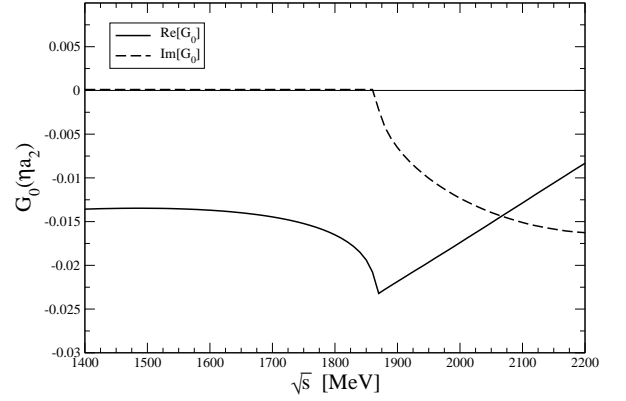}
\caption{$G_0$ function for the $\eta a_2$ channel.}
\label{fig:G0}
\end{center}
\end{figure}

On the other hand, taking into account the general form
of the $S$-matrix of an
 $AB$ interaction 
\ba
S&=&-i T(2\pi)^4 \delta(k+k_B-k'-k'_B)\frac{1}{{\cal V}^2}\nn\\
&&\times\frac{1}{\sqrt{2 \omega_k}} 
\frac{1}{\sqrt{2 \omega_{k'}}}
\frac{1}{\sqrt{2 \omega_{k_B}}}
\frac{1}{\sqrt{2 \omega_{k'_B}}}.
\label{eq:finalS2AB}
\ea
and comparing this equation with Eqs.~(\ref{eq:single}) and 
(\ref{eq:finalS2}), we 
obtain that  the 
FCA equations (\ref{eq:faddeev})  take in our case the form
\begin{eqnarray}
T_{Ab_1}&=&  F^{(s)}_{B,1}\bar t_1 + \bar t_1 G_0 T_{Ab_2}  \nonumber\\
T_{Ab_2}&=& F^{(s)}_{B,2}\bar t_2  + \bar t_2 G_0 T_{Ab_1}\nonumber\\
T&=&T_{Ab_1}+T_{Ab_2}
\label{eq:Faddeevsystem}
\end{eqnarray}
with
\be \bar t_i=  \sqrt{\frac{\omega_{k_B} \omega_{k_B'}}{\omega_{p_i}
 \omega_{p_i'}}}
t_{A b_i}(s_i).
\ee
Note that the argument of the  $t_{A b_i}$ function is $s_i$ of
 Eq.~(\ref{eq:si}), whereas the total amplitude $T$ can be
regarded as a function of the global $s$. 
In Eq.~(\ref{eq:Faddeevsystem})
 \ba
\label{eq:G0}
&&G_0(s;A,B)=\\ \nn
&&\frac{1}{2 \sqrt{\omega_{k_B}\omega_{k_B'}}}
\int\frac{d^3q}{(2\pi)^3} 
F_B(\vec q)
\frac{1}{{q^0(s;A,B)}^2-\vec{q}\,^2-m_A^2+i\epsilon}.
\ea
As an example, we show in Fig.~\ref{fig:G0}
 the real and imaginary parts of the
$G_0$ function
for the channel $\eta a_2$. Close to the threshold it resembles the typical
shape
of the loop function of two mesons, in this case one $\eta$ and one $a_2$.
 For the other channels $G_0$
has similar qualitative shapes but with different thresholds.

Solving algebraically the system of equations~(\ref{eq:Faddeevsystem})
 gives the following final three-body scattering
amplitude

\ba
T&=&T_{Ab_1}+T_{Ab_2}=\frac{\bar t_1+\bar t_2+2\bar t_1 
\bar t_2  G_0}
{1-\bar t_1 \bar t_2 \widetilde G_0^2}\nn\\
&&+\bar t_1 (F^{(s)}_{B,1}-1)+\bar t_2 (F^{(s)}_{B,2}-1)
\ea

Thus far we have not considered anywhere the finite width of the
tensor resonance $B$. We have taken this effect into account by folding the
final amplitude $T(M_B)$, regarded as a function of  $M_B$,
with the spectral function of the $B$ resonance:

\ba
\label{eq:convol}
&&T\to
T=\frac{1}{\cal N}_B
\int_{(M_B-2\Gamma^0_B)^2}^{(M_B+2\Gamma^0_B)^2}
ds_B\,\\ \nonumber
&&\times Im
\left\{\frac{1}{s_B-M_B^2+iM_B\Gamma_B(s_B)}\right\}
T(\sqrt{s_V}),
\ea
\ba
{\cal N}_B=
\int_{(M_B-2\Gamma^0_B)^2}^{(M_B+2\Gamma^0_B)^2}
ds_B\,
Im
\left\{\frac{1}{s_B-M_B^2+iM_B\Gamma_B(s_B)}\right\}.\nn
\ea
where 
$\Gamma_B(s_B)$ is the energy dependent width of the $B$ particle
and $\Gamma^0_B=\Gamma_B(M_B^2)$.


Thus far we have only considered the interaction of one 
single channel consisting of one
pseudoscalar meson and one tensor meson. We are going to estimate the
possible coupled channels effect. If we look
 at
table~\ref{tab:tsI12}
in the Appendix,
 we see that for total isospin $I=1/2$ of the $PVV$ system we can
have for single scattering interaction non diagonal scattering amplitudes.
(This is not the case for $I=1$  and $I=0$).
For instance, an initial $\pi K^*_2$ state can turn into
$K a_2$ thanks to
the transition $t_{\pi\rho,K \bar K^*}$. However a
 direct application of the
FCA cannot be done in this case 
since the FCA requires that the $B$ cluster, the tensor meson in our case,
 is not much
altered by the interaction with the $A$ particle, the pseudoscalar meson.
 Otherwise one should evaluate the mechanisms of
the multistep processes with quantum field theory evaluating the
corresponding three meson loops, etc. This will spoil the
simplicity of the FCA approximation since
the problem becomes very involved with the higher iterations
 and ultimately turns out into a
problem far more complicated than the use of the exact Faddeev equations
 from the beginning. 
The wave functions used in the derivation of the
S-matrix and the form factors (see Ref.~\cite{Roca:2010tf})
 are now different for the initial
and final cluster
$B$ and thus the derivation in this section is not directly valid.
The differences are essentially due to the different masses between
different
channels. However, in the present case
the constituent particles of the $B$ resonance
of the initial and final state are all vector mesons, and 
thus they
have a similar mass and so are the typical momenta inside the clusters $B$.
On the other hand, doing coupled channels is in general relevant if the
final amplitude for 
the different channels have a similar strength. However, advancing some
results, this is not the case in the present work. From all this reasons we
can conclude that the couple channel effects will be small and thus we can
just estimate its effect adapting the formalism discussed so far.

The equation for the coupled channel estimation is now formally 
the same than 
Eq.~(\ref{eq:Faddeevsystem}) but now the amplitudes are $4\times 4$
matrices
(since we have four pseudoscalar-tensor channels for $I=1/2$) such that 
$(T^i)_{jk}$, $i=1,2$; $j,k=1-4$, represents the interaction 
$T^i_{{A_j b_{i,j}}\to A_k b_{i,k}}$ of channel $j$ starting with
particle $i$ of the cluster to produce channel $k$.
Now $(\bar t^i)_{jk}=\sqrt{\frac{\omega_{B_j} \omega_{B_k}}
{\omega_{i,j} \omega_{i,k}}}
t^i_{{A_j b_{i,j}}\to A_k b_{i,k}}$. The $G_0$ function is now a diagonal
$4\times 4$ matrix which elements are $G_0(s;A_i,B_i)$, $i=1,4$.
The form factor that multiplies the single scattering contribution in 
Eq.~(\ref{eq:Faddeevsystem}) is now a diagonal matrix which element $jj$ is
$F^{(s)}_{B_j,i}$.  
Finally, to take into account the width of the $B$ particles, 
an independent folding with their spectral functions 
analogous to Eq.~(\ref{eq:convol}) is implemented 
for every different $B$ species.

\section{Results}
\label{sec:results}
\begin{figure}[!ht]
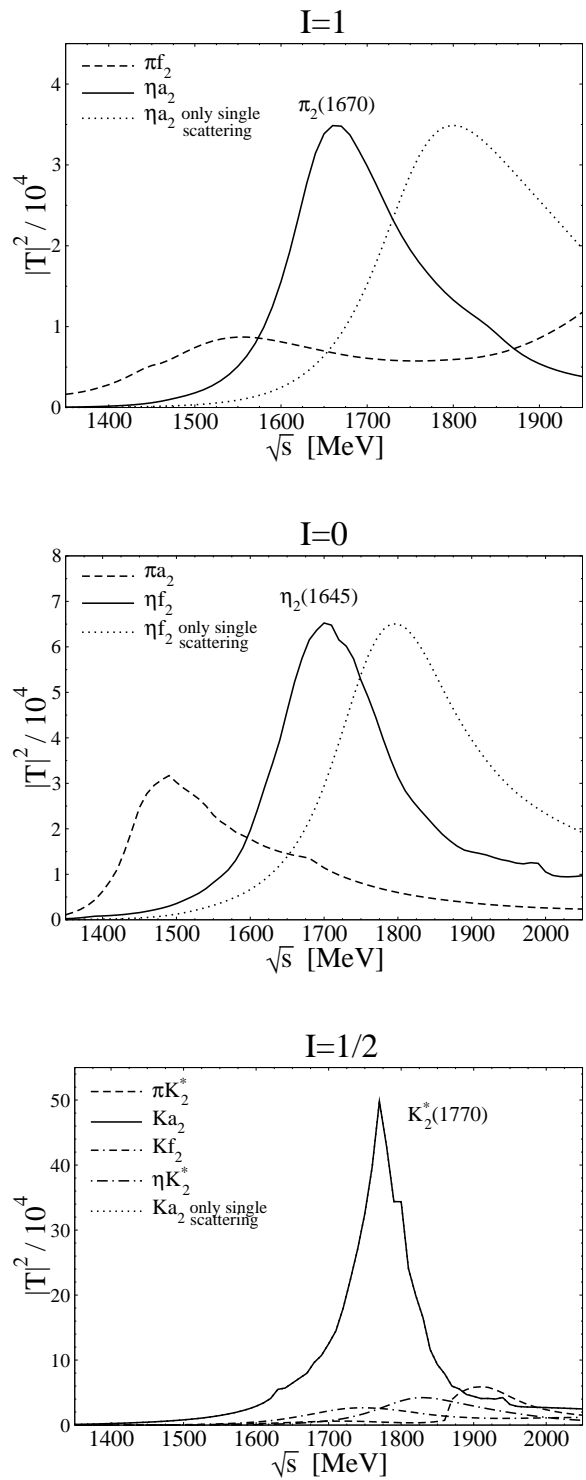

     \centering
      \subfigure[]{
          \includegraphics[width=.88\linewidth]{figure7a.eps}} \\
      \subfigure[]{
          \includegraphics[width=.88\linewidth]{figure7b.eps}} \\
       \subfigure[]{
          \includegraphics[width=.88\linewidth]{figure7c.eps}}
   \caption{ Modulus squared of the three-body 
pseudoscalar-vector-vector
scattering amplitudes for the different total isospin channels. 
No coupled channels effect is considered for the $I=1/2$ case.
The particle labels over the dominant peaks
indicate the experimental pseudotensor $2^{-+}$ mesons 
to which our dynamically generated are associated.}
     \label{fig:PVVamplitudes}
\end{figure}

In Fig.~\ref{fig:PVVamplitudes} we show the modulus squared of the  three-body
scattering amplitudes for each channel and for total isospin $I=1$,
$I=0$ and $I=1/2$. The calculation accounts for the full model but
without coupled channels in $I=1/2$, which will be discussed later on.
In all the plots the dotted line represent the results for the dominant
channels but considering only  the single scattering contribution. The
single scattering  plots have been normalized to the peak of the
dominant channel for every isospin in order to make easier the
comparison of the position of the maxima (the real size of the 
dotted plots for $I=1$ and $I=0$ 
is actually about a factor three larger than the
solid lines). In the $I=1/2$ case the dotted and solid lines overlap.
The particle labels over the dominant peaks
indicate the experimental pseudotensor mesons 
to which we associate our dynamically generated resonances. 

Considering only the single scattering, resonant shapes are
clearly visible in the plots but the position of the maxima do not agree
well with the experimental value of the closest resonance in the
corresponding channel, except for $I=1/2$.
 In table~\ref{tab:masses} the value
of the masses of the dynamically generated pseudotensor systems are
shown in comparison with the experimental values at the PDG
\cite{pdg}.
\begin{table*}[!t]
\begin{center}
\begin{tabular}{|c|c|c|c|c|} 
\hline
  \begin{tabular}{c} assigned \\ resonance \end{tabular} &
  \begin{tabular}{c} dominant \\ channel \end{tabular} & 
  \begin{tabular}{c} mass \\ PDG \cite{pdg} \end{tabular}  
& \begin{tabular}{c}mass, only \\ single scatt. \end{tabular} & \begin{tabular}{c}mass \\ full model \end{tabular}  \\ \hline
$\pi_2(1670$   & $\eta a_2(1320)$   & $1672\pm 3$   & 1800    & 1660  \\ \hline
$\eta_2(1645)$ & $\eta f_2(1270)$   & $1617\pm 5$ & 1795    & 1695   \\ \hline
$K^*_2(1770)$  & $K a_2(1320)$      & $1773\pm 8$ & 1775    & 1775   \\ \hline
 \end{tabular}
\end{center}
\caption{Results for the masses of the dynamically generated pseudotensor resonances. (All units are MeV)}
\label{tab:masses}
\end{table*}
The second column of table~\ref{tab:masses} indicates the dominant
channel, which is the one chosen to get the
mass quoted in the table obtained from the position of the maximum.

 When  the multiple scattering mechanisms are added, an important
improvement in the agreement with experimental masses is obtained for
$I=1$ and $I=0$.  For the $I=1/2$ channel no change is appreciable when
adding the multiple scattering mechanisms but for this channel the mass
obtained with  single scattering contribution already agreed well with
the experimental value for the $K^*_2(1770)$ resonance. The improvement
obtained when considering the full mechanisms is an indication of the
goodness and validity of the model proposed in the present work
for the nature of the pseudotensor mesons considered.
From the width of the
amplitudes squared we can estimate the width of the dynamically
generated pseudotensor states.
We get for the $\pi_2(1670)$, $\eta_2(1645)$ and $K^*_2(1770)$,
160~MeV, 170~MeV and 80~MeV respectively, to be compared with
 the experimental
values $260\pm 9$, $181\pm 11$ and $186\pm 14$ respectively.
The underestimation of the width within our model is
not worrying. 
 Indeed it is expected because
a good reproduction of the width would imply to account also for
 other components and possible decay channels which can
contribute to the decay width even if they do not significantly affect
the mass. Therefore, 
in contrast with the mass results, which are quite reliable,
our results for the width must be considered only qualitatively.

It is worth stressing the simplicity of our approach
and that  there are no free parameters in
the model for the three-body scattering,
once the regularization parameters of
the $VV$ and $PV$ amplitudes are slightly changed to agree better
with the tensor and axial-vector experimental masses respectively, as
explained in section~\ref{sec:2}. Therefore the results and
conclusions of the present work are genuine predictions with no fits
to any pseudotensor meson parameter or experimental data.

The results allow us to conclude that 
our three-body model generates
dynamically the $\pi_2(1670)$ as a dominantly $\eta a_2$ molecule,
the $\eta_2(1645)$ as $\eta f_2$ and the $K^*_2(1770)$ as $K a_2$.
Certainly other Fock space components like quark-antiquark,
two meson components (different to those considered in the present
work), etc, are present in the pseudotensor resonances.
However the fact that with the picture of a pseudoscalar-tensor
molecule, with the tensor itself being two vector mesons,
we get good agreement for the value of the masses reinforces the idea
that this component is the dominant one in the wave function of these
pseudotensor mesons.
The extra components could only affect the total width of
the resonances, as explained above.
 
 \begin{figure}[!h]
\begin{center}
\includegraphics[width=0.88\linewidth]{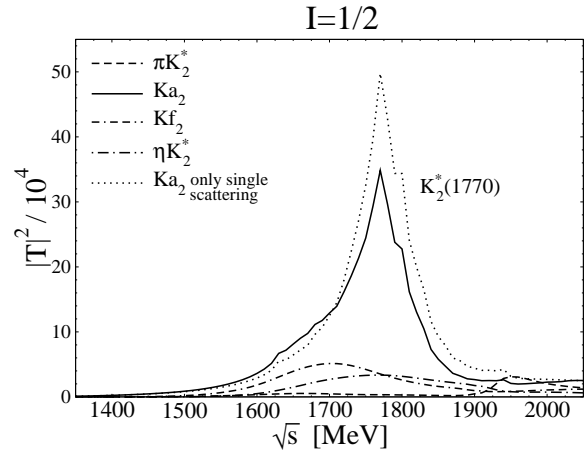}
\caption{Same meaning as in Fig.~\ref{fig:PVVamplitudes}
 for the $I=1/2$ 
case but
implementing coupled channels.}
\label{fig:T2I12coupled}
\end{center}
\end{figure}
In Fig.~\ref{fig:T2I12coupled}
 we show the coupled channel effect in the $I=1/2$
channel. We see, by comparing with Fig.~\ref{fig:PVVamplitudes} that the 
effect is very small and
in this particular case, since there is one channel so strongly
dominant, $K a_2$, the coupled channel effects is negligible and
thus there
is no need to improve upon the estimation done in the present work.

\section{summary}
\label{sec:concl}

We have performed a theoretical study of the three body system
consisting of one pseudoscalar and two vector mesons were the vector
mesons are strongly correlated forming a tensor resonance.
The motivation was that in previous works it was obtained that two
vector mesons in spin 2 and s-wave tend to bind making up the lightest
tensor mesons $2^{++}$ and the interaction of a pseudoscalar and a
vector meson in s-wave is also very attractive and generate dynamically
the lightest axialvector resonances. Thus the $PVV$ system could be
strongly attractive and generate $2^{-+}$ resonant states.

The three body amplitudes are evaluated solving the Faddeev
equations in the fixed center approximation, which can be applied since
two of the three particles are clustered.
The three body amplitudes are written in terms of the unitarized $PV$
interactions, which are obtained from the application of the techniques
of the chiral unitary approach.
The method used for the three-body evaluation 
does not introduce any new parameter
once the regularization in the $VP$ and $VV$ is chosen.
This allows us to make genuine predictions which can be compared with 
experimental values for pseudotensor resonances.

In the three body amplitudes we obtain significant resonant signals
which can be
associated with the $\pi_2(1670)$, $\eta_2(1645)$ and $K^*_2(1770)$
experimental pseudotensor resonances, where the dominant channels
 in the making
up of this resonances are $\eta a_2(1320)$, 
$\eta f_2(1270)$ and  $K a_2(1320)$ respectively.

In spite that other states like quark-antiquark,
other meson-meson,
several mesons, etc, can contribute to the wave functions
of these pseudotensor resonances,
the remarkable agreement obtained with our picture
make us to consider the pseudoscalar-tensor mesons contribution
as the dominant component in the building up of these resonances.

\section*{Acknowledgments}
We thank E.~Oset for useful discussions.
This work is partly supported by DGICYT contracts  FIS2006-03438,
FPA2010-17806, the Fundaci\'on S\'eneca grant 11871/PI/09 and
the EU Integrated Infrastructure Initiative Hadron Physics
Project  under Grant Agreement n.227431.

\section*{Appendix: Single scattering in terms of two-body amplitudes}
\label{sec:appendixA}

We derive in this Appendix
the exact expression of the single scattering amplitudes, 
$t_i$, in terms of the unitarized
$PV$ scattering amplitudes. The latter
are calculated for a given $PV$ isospin. Therefore, we need
to write the total isospin state of the global system $AB$, 
$| I_A,I_B,I,M  \rangle$, 
 in terms of the coupled
isospin state of particle $A$ and $b_i$, 
$|I_A,I_i,I_{Ai},M_{Ai}\rangle\otimes|I_j, M_j\rangle$:

\ba
& &| I_A,I_B,I,M  \rangle^{(i)} =\nn \\
& &\sum_{I_{Ai}}\sum_{M_{Ai}}\sum_{M_A}
{\cal C}(I_A,I_B,I|M_A,M-M_A,M)\nn \\ 
&&\times
{\cal C}(I_i,I_j,I_B|M_{Ai}-M_A,M-M_{Ai},M-M_A)\nn \\ 
&&\times
{\cal C}(I_A,I_i,I_{Ai}|M_A,M_{Ai}-M_A,M_{Ai})[\eta-\delta_{i1}
(\eta-1)]\nn \\ 
&&\times|I_A,I_i,I_{Ai},M_{Ai}\rangle\otimes|I_j, M-M_{Ai}\rangle
\label{eq:stateclebsh}
\ea 
where the $i$ label means that we are correlating the particle $A$ with
particle $b_i$, $i=1,2$, $j\neq i$, 
$I_A$ is the isospin of the particle $A$, $I_B$ the isospin of particle
$B$, $I$ the total $AB$ isospin, $I_i$ the isospin of particle $b_i$,
 $I_{Ai}$ the global isospin of the $A$-$b_i$ system,
  and the $M_x$ are the third components of the corresponding
isospins.   In Eq.~(\ref{eq:stateclebsh}), $\eta=(-1)^{I_1+I_2-I_B}$,
$\delta$ is the Kronecker delta and 
${\cal C}(j_1,j_2,j_3|m_1,m_2,m_3)$ represent Clebsch-Gordan coefficients.
For example, for total isospin $I=0$, one of the possible channels
is  $\pi a_2$, with $A=\pi$,
$B=a_2$, $b_1=K^*$ and $b_2=\bar K^*$. In this case for the $A b_1$
interaction we
 have
\ba
( \pi a_2  )^{(1)}_{I=0,M=0}&=&-\frac{1}{\sqrt{2}}
(\pi K^*)_{1/2,-1/2}\bar K^{*0}\nn\\
&&-\frac{1}{\sqrt{2}}(\pi K^*)_{1/2,+1/2}K^{*-}.
\ea
Throughout the work we have used the following isospin
conventions: 
$|\pi^+\rangle=-|1,+1\rangle$,
$|\rho^+\rangle=-|1,+1\rangle$,
$|a_2^+\rangle=-|1,+1\rangle$,
$|K^-\rangle=-|\frac{1}{2},-\frac{1}{2}\rangle$,
$|{K^*}^-\rangle=-|\frac{1}{2},-\frac{1}{2}\rangle$,
$|{K^*}_2^-\rangle=-|\frac{1}{2},-\frac{1}{2}\rangle$
(for the other particles the sign is positive)
as is usually used in chiral perturbation theory and in the work
from which our $PV$ amplitudes is based \cite{Roca:2005nm}.

The scattering potential for the single scattering contribution can be
written in terms of the two body amplitudes, $t_{Ab_i,A'b_{j'}}$,
for the transition $Ab_i\to A'b_{j'}$:

\ba
&&^{(i)}\langle Ab_1 b_2 | V | A' b'_1 b'_2 \rangle^{(i')}
=\sum_{I_{Ai}}\bigg[\sum_{M_{Ai}}\sum_{M_A}\sum_{M_A'}\nn \\ 
&&\times
{\cal C}(I_A,I_B,I|M_A,M-M_A,M)\nn \\ 
&&\times{\cal C}(I_{A'},I_{B'},I|M_{A'},M-M_{A'},M)\nn \\ 
&&\times
{\cal C}(I_i,I_j,I_B|M_{Ai}-M_A,M-M_{Ai},M-M_A)\nn \\ 
&&\times
{\cal C}(I_{i'},I_{j'},I_{B'}|M_{Ai'}-M_A,M-M_{Ai'},M-M_{A'})\nn \\ 
&& \times
{\cal C}(I_A,I_i,I_{Ai}|M_A,M_{Ai}-M_A,M_{Ai})\nn \\ 
&& \times{\cal C}(I_{A'},I_{i'},I_{Ai}|M_{A'},M_{Ai}-M_{A'},M_{Ai})\bigg]
\nn \\ 
&& \times
[\eta-\delta_{i1}(\eta-1)][\eta'-\delta_{i'1}(\eta'-1)]
\ t^{(I_{Ai})}_{Ab_i,A'b_{i'}}\nn \\ 
& \equiv& \sum_{I_{Ai}} \alpha_{i,i'}\ t^{(I_{Ai})}_{Ab_i,A'b_{i'}}.
\label{eq:V121p2p}
\ea
Note that only 
the diagonal $t_{Ab_i,Ab_i}$ is
needed if not doing coupled channels. 
The expression in Eq.~(\ref{eq:V121p2p}) is very convenient
and easily implementable
 for computer evaluation and is general for
any three-body system made of $A$, $b_1$ and $b_2$ and thus can be
useful for
other works where the FCA is applied.
In tables~\ref{tab:tsI1}, \ref{tab:tsI0} and  \ref{tab:tsI12}
 we show the two-body amplitudes obtained from
Eq.~(\ref{eq:V121p2p}) for the three different total $PVV$ isospins.
The different $PVV$ channels are labeled by the pseudoscalar meson, the
tensor resonance and in brackets the two vector mesons dominant in the
formation of the tensor meson.
For $I=1$ and $I=0$ the non-diagonal terms are zero and thus what we show
in the table represent diagonal elements. For $I=1/2$ the first column
represent the initial channel and the first row the final one. 
The non-diagonal terms
are only needed in the estimation of the coupled channels.

\begin{table}[h]
\begin{center}
\begin{tabular}{|l|c|c|} 
\hline
        & $\pi f_2(\rho \rho)$  & $\eta a_2(K^* \bar K^*)$     \\ \hline 
$t_1$   &  $\frac{2}{9} t^{(I=0)}_{\pi\rho,\pi\rho} + \frac{2}{3} t^{(I=1)}_{\pi\rho,\pi\rho}+\frac{10}{9} t^{(I=2)}_{\pi\rho,\pi\rho}$  &	$t^{(I=1/2)}_{\eta K^*,\eta K^*}$  \\ \hline
$t_2$   &  $\frac{2}{9} t^{(I=0)}_{\pi\rho,\pi\rho} + \frac{2}{3} t^{(I=1)}_{\pi\rho,\pi\rho}+\frac{10}{9} t^{(I=2)}_{\pi\rho,\pi\rho}$	&     	$t^{(I=1/2)}_{\eta K^*,\eta K^*}$  \\ \hline
 \end{tabular}
\end{center}
\caption{Three body single scattering amplitudes in terms of the 
unitarized two-body ($PV$) amplitudes for total isospin $I=1$}
\label{tab:tsI1}
\end{table}

\begin{table}[h]
\begin{center}
\begin{tabular}{|l|c|c|} 
\hline
        & $\pi a_2(K^* \bar K^*)$  & $\eta f_2(\rho \rho)$     \\ \hline 
$t_1$   & $ t^{(I=1/2)}_{\pi K^*,\pi K^*}$  & $2 t^{(I=1)}_{\eta\rho,\eta\rho}$	  \\ \hline
$t_2$   & $ t^{(I=1/2)}_{\pi K^*,\pi K^*}$  & $2 t^{(I=1)}_{\eta\rho,\eta\rho}$ \\ \hline
 \end{tabular}
\end{center}
\caption{Three body single scattering amplitudes in terms of the 
unitarized two-body ($PV$) amplitudes for total isospin $I=0$}
\label{tab:tsI0}
\end{table}

\begin{table*}[ht]\tiny
\begin{center}
\begin{tabular}{|l|c|c|c|c|} 
\hline
                        & $\pi K^*_2(\rho K^*)$  & $K a_2(K^* \bar K^*)$   & $K f_2(\rho \rho)$       & $\eta K^*_2(\rho K^*)$  \\ \hline 
$\pi K^*_2(\rho K^*)$   &  \begin{tabular}{l} $t_1= \frac{1}{3} t^{(I=0)}_{\pi\rho,\pi\rho} + \frac{2}{3} t^{(I=1)}_{\pi\rho,\pi\rho}$  	      \\ 
                           		      $t_2= \frac{1}{9} t^{(I=1/2)}_{\pi K^*,\pi K^*} + \frac{8}{9} t^{(I=3/2)}_{\pi K^*,\pi K^*}$	      \end{tabular}		  
			&  \begin{tabular}{l} $t_1= -\frac{1}{2} t^{(I=0)}_{\pi\rho,K \bar K^*} + \frac{1}{\sqrt{6}} t^{(I=1)}_{\pi\rho,K \bar K^*}$  \\
			   		      $t_2=0$												      \end{tabular} 
			&  \begin{tabular}{l} $t_1=0$												      \\
			   		      $t_2= -\frac{2}{3\sqrt{3}} t^{(I=1/2)}_{\pi K^*,K\rho}-\frac{8}{3\sqrt{3}} t^{(I=3/2)}_{\pi K^*,K\rho}$ \end{tabular}  
			&  \begin{tabular}{l} $t_1=\sqrt{\frac{2}{3}}t^{(I=1)}_{\eta\rho,\eta\rho}$						      \\
				 	      $t_2= -\frac{1}{3} t^{(I=1/2)}_{\pi K^*,\eta  K^*}$						      \end{tabular}   \\ \hline
$K a_2(K^* \bar K^*)$   &  \begin{tabular}{l} $t_1= 0$  											      \\ 
                           		      $t_2=  -\frac{1}{2} t^{(I=0)}_{K \bar K^*,\pi\rho} + \frac{1}{\sqrt{6}} t^{(I=1)}_{K \bar K^*,\pi\rho}$ \end{tabular}		  
			&  \begin{tabular}{l} $t_1= 0 $ 											      \\
			   		      $t_2=\frac{3}{4} t^{(I=0)}_{K \bar K^*,K \bar K^*} + \frac{1}{4} t^{(I=1)}_{K \bar K^*,K \bar K^*}$     \end{tabular} 
			&  \begin{tabular}{l} $t_1=0$												      \\
			   		      $t_2= 0$  											      \end{tabular}  
			&  \begin{tabular}{l} $t_1=0$												      \\
					      $t_2= \frac{1}{2} t^{(I=1)}_{K \bar K^*,\eta\rho}$                                                      \end{tabular}   \\ \hline
$Kf_2(\rho \rho)$       &  \begin{tabular}{l} $t_1= -\frac{1}{3\sqrt{3}} t^{(I=1/2)}_{K\rho,\pi K^*}  - \frac{4}{3\sqrt{3}} t^{(I=3/2)}_{K\rho,\pi K^*}$ \\ 
                           		      $t_2= -\frac{1}{3\sqrt{3}} t^{(I=1/2)}_{K\rho,\pi K^*}  - \frac{4}{3\sqrt{3}} t^{(I=3/2)}_{K\rho,\pi K^*}$ \end{tabular}		   
			&  \begin{tabular}{l} $t_1= 0 $		   									                 \\
			   		      $t_2=0$                                                                                                    \end{tabular} 
			&  \begin{tabular}{l} $t_1=\frac{2}{3} t^{(I=1/2)}_{K\rho,K\rho} + \frac{4}{3} t^{(I=3/2)}_{K\rho,K\rho}$                        \\
			   		      $t_2=\frac{2}{3} t^{(I=1/2)}_{K\rho,K\rho} + \frac{4}{3} t^{(I=3/2)}_{K\rho,K\rho}$ 	                 \end{tabular}  
			&  \begin{tabular}{l} $t_1=\frac{1}{\sqrt{3}} t^{(I=1/2)}_{K\rho,\eta K^*}$  							 \\
					      $t_2= \frac{1}{\sqrt{3}} t^{(I=1/2)}_{K\rho,\eta K^*}$                                                     \end{tabular}   \\ \hline
$\eta K^*_2(\rho K^*)$  &  \begin{tabular}{l} $t_1= \sqrt{\frac{2}{3}}t^{(I=1)}_{\eta\rho,\pi\rho}$                                                     \\ 
                           		      $t_2=  -\frac{1}{3} t^{(I=1/2)}_{\eta K^*,\pi  K^*}$                                                       \end{tabular}		   
         		&  \begin{tabular}{l} $t_1= \frac{1}{2} t^{(I=1)}_{K \eta\rho,K\bar K^*} $		   					 \\
			   		      $t_2=0$                                                                                                    \end{tabular} 
			&  \begin{tabular}{l} $t_1=0$                                                                                                    \\
			   		      $t_2=\frac{2}{\sqrt{3}} t^{(I=1/2)}_{\eta K^*,K\rho}$              	                                 \end{tabular}  
			&  \begin{tabular}{l} $t_1=t^{(I=1)}_{\eta \rho,\eta\rho}$  							                 \\
					      $t_2= t^{(I=1/2)}_{\eta K^*,\eta K^*}$                                                     \end{tabular}   \\ \hline
 \end{tabular}
\end{center}
\caption{ Three body single scattering amplitudes in terms of the 
unitarized two-body ($PV$) amplitudes for total isospin $I=1/2$}
\label{tab:tsI12}
\end{table*}


At this point let us explicit further the explanation why the  $K \bar
K^*_2$ and $\bar K K^*_2$ channels do not contribute to the $I=1$ and
$I=0$ three-body amplitudes. We will discuss the $I=1$ case since the $I=0$
reasoning is totally analogous. The channel with $I=1$ that we are
considering has negative $G$-parity.  The $K \bar K^*_2$ and
 $\bar K K^*_2$ states
do not have defined $G$-parity by themselves. However if we define
$KK^*_2(\pm)\equiv  1/\sqrt{2}|K \bar K^*_2 \mp \bar K K^*_2\rangle$,
it  is eigenstate of $G$ with eigenvalue $\pm1$.  Thus in principle we
should include $KK^*_2(-)$ as another channel in the global $I=1$ case.
 Let us consider first
the $|K \bar K^*_2\rangle$ channel in $I=1$, $M=+1$. For the evaluations
needed in the present work,
we take the $K^*_2$ as a $\rho K^*$ state and for the evaluation of
the FCA we need the interaction of the $K$ with the $\rho$ and the
$\bar K^*$. Let us consider first the interaction of the kaon with
the $\bar K^*$.
Applying Eq.~(\ref{eq:stateclebsh}) we have
\ba
&&(K \bar K^*_2)_{1,+1}\longrightarrow\nn\\
&&-\frac{1}{\sqrt{3}} \Big[
(K \bar K^{*})_{0,0}\rho^+
- (K \bar K^{*})_{1,0}\rho^+
- (K \bar K^{*})_{1,+1}\rho^0\Big]\ ,\nn\\
&&(\bar K K^*_2)_{1,+1}\longrightarrow\nn\\
&&-\frac{1}{\sqrt{3}} \Big[
(\bar K K^{*})_{0,0}\rho^+
+ (\bar K  K^{*})_{1,0}\rho^+
 +(\bar K K^{*})_{1,+1}\rho^0\Big]
\ea
Therefore we have to do
\ba
KK^*_2(-)&=&\frac{1}{\sqrt{2}}(K \bar K^*_2 + \bar K K^*_2
)_{1,+1}\nn\\
\longrightarrow&&-\frac{1}{\sqrt{3}} \Big[
\frac{1}{\sqrt{2}}(K \bar K^{*}+\bar K  K^{*} )_{0,0}
\rho^+\nn\\
&&
- \frac{1}{\sqrt{2}}(K \bar K^{*}+\bar K  K^{*} )
_{1,0}\rho^+\nn\\
&&
-
 \frac{1}{\sqrt{2}}(K \bar K^{*}+\bar K  K^{*} )_{1,+1}\rho^0\Big].
 \label{eq:Gcomb}
\ea

But $ \frac{1}{\sqrt{2}}(K \bar K^{*}+\bar K  K^{*} )$ is eigenstate of
$G$ with eigenvalue $+1$ and, on the other hand, 
 $\rho$ has $G$-parity $(+)$.  Therefore
under a $G$-parity transformation the right member of 
Eq.~(\ref{eq:Gcomb}) has $G$-parity $(+)$ which is of opposite sign
 to what is
required by the left term of Eq.~(\ref{eq:Gcomb}). Therefore this
channels is not permitted in the evaluation of the FCA of the Faddeev
equations. An analogous reasoning leads to the same conclusion for the
interaction of the kaon with the $\rho$ and also for the interaction
needed in the global $I=0$ channel.

\end{document}